# A cyanine dye rotaxane porphyrin nanoring complex as a model light harvesting system

Jiratheep Pruchyathamkorn,[a] William J. Kendrick,[a] Andrew T. Frawley,[a] Andrea Mattioni,[b] Felipe Caycedo-Soler,[b] Susana F. Huelga,[b] Martin B. Plenio,*[b] and Harry L. Anderson*[a]

**Abstract:** A nanoring-rotaxane supramolecular assembly, with a Cy7 cyanine dye (hexamethylindotricarbocyanine) threaded along the axis of the nanoring, has been synthesized as a model for the energy transfer between the light harvesting complex LH1 and the reaction center in purple bacteria photosynthesis. The complex displays efficient energy transfer from the central cyanine dye to the surrounding zinc porphyrin nanoring. We present a theoretical model that reproduces the absorption spectrum of the nanoring and quantifies the excitonic coupling between the nanoring and the central dye, explaining the efficient energy transfer and elucidating the similarity with structurally related natural light harvesting systems.

Living organisms achieve photosynthesis using a wide diversity of supramolecular chlorophyll arrays to absorb sunlight and channel energy to a reaction center (RC), where it is converted to chemical potential.[1] The design requirements for an efficient light-harvesting system are poorly understood, despite many investigations of natural and artificial chromophore arrays.[2,3] All species of purple bacteria use an assembly known as light-harvesting complex 1 (LH1), which surrounds the RC. There is some variation in the structures of the LH1-RC complexes from different species of bacteria, but they typically consist of a ring of about 32 bacteriochlorophyll molecules encircling the RC.[4] The energy transfer process from LH1 to the RC is intriguing because it occurs rapidly (within about 50 ps) despite being reversible and thermodynamically up-hill; back transfer from the RC to LH1 is even faster, occurring within a few picoseconds.[5] Several artificial systems have been synthesized as models of the LH1-RC supercomplex, to investigate energy transfer between a ring of chromophores and a central acceptor/donor chromophore.[3,6] Here, we present the design and synthesis of a supramolecular dye complex inspired by the LH1-RC system, and examine its photophysical properties.

Porphyrin nanorings offer a rigid cyclic array of absorbing chromophores, resembling the ring of bacteriochlorophyll pigments in LH1. These nanorings are synthesized using templates of the correct geometry to bind the metal centers of each metalloporphyrin unit. A variety of oligopyridine ligands can be used as templates,[7] including pyridine-functionalized cyclodextrins (CDs).[8] Oligopyridine templates bind strongly to the nanorings, with association constants ($K_f$) in the range $10^{29}$–$10^{36}$ M$^{-1}$.[9] These compounds provide a unique toolkit with which to mimic light-harvesting systems.[10]

In order to model the LH1-RC supercomplex, a chromophore must be positioned at the center of the porphyrin nanoring, to mimic the RC. α-CD provides an ideal scaffolding for achieving this spatial arrangement: its six primary OH groups can be functionalized with 4-pyridyl substituents to form a template for a six-porphyrin nanoring (*c*-P6), while its hydrophobic cavity can be used to encapsulate guest molecules, such as the hexamethylindotricarbocyanine dye (HITC, **Cy7**), via rotaxane formation.[11] This concept led us to target the structure **Cy7⊂*c*-P6·T6*** (Figure 1 and Scheme 1): the *c*-P6 and **Cy7** units mimic the LH1 and RC respectively, and the α-CD fixes the spatial arrangement of the chromophores. Molecular mechanics calculations predicted two low-energy conformations for **Cy7⊂*c*-P6·T6***, labeled as **A** and **B** in Figure 1. Both conformations have the cyanine dye aligned along the axis of the porphyrin nanoring. In conformation **A**, the center of the cyanine is near the plane of the six zinc centers, whereas in **B**, the **Cy7⊂T6*** unit is shifted towards one rim of the nanoring. The calculated energies of these conformations almost the same (**A** is lower in energy by 0.7 kJ mol$^{-1}$); we expect them both to be populated and to interconvert rapidly in solution.

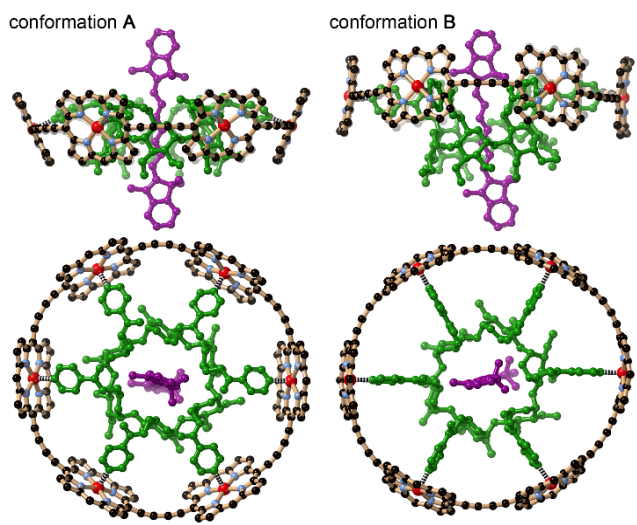

**Figure 1.** Two low-energy conformations of **Cy7⊂*c*-P6·T6*** from molecular mechanics calculations; two orthogonal views of each conformation (MM+ force field, *meso*-aryl groups and hydrogen atoms are omitted for clarity, see SI Section 3 for details).

[a]  J. Pruchyathamkorn, Dr. W. J. Kendrick, Dr. A. T. Frawley, Prof. H. L. Anderson
    Department of Chemistry, Oxford University
    Chemistry Research Laboratory, Oxford, OX1 3TA, UK
    E-mail: harry.anderson@chem.ox.ac.uk
[b] A. Mattioni, Dr. F. Caycedo-Soler, Prof. S. F. Huelga, Prof. M. B. Plenio
    Institute of Theoretical Physics and IQST
    Ulm University
    Albert-Einstein-Allee 11, 89069 Ulm, Germany
    Email: martin.plenio@uni-ulm.de

Supporting information for this article is given via a link at the end of the document.

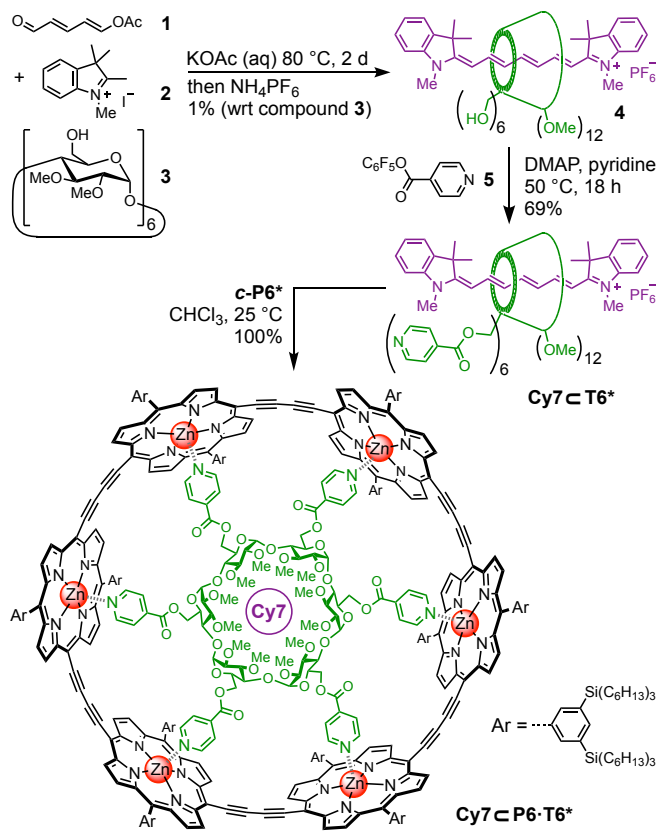

**Scheme 1.** Synthesis of **Cy7⊂c-P6·T6*** (cyanine dye: purple; functionalized CD: green).

The synthesis of **Cy7⊂c-P6·T6*** was achieved as summarized in Scheme 1. The first step is the reaction of 5-oxopenta-1,3-dienyl acetate **1**, tetramethyl indolium iodide **2** and per-2,3-di-O-methyl-α-cyclodextrin[12] **3** in aqueous potassium acetate to form rotaxane **4**. The yield of this rotaxane is low (1%, based on **3**) because the reaction needs to be carried out in water, and it is not practical to use a large excess of the methylated cyclodextrin, as would be done with native α-CD,[11] but sufficient material can be prepared to complete the synthesis. Acylation of all six hydroxyl groups of rotaxane **4** with isonicotinic acid pentafluorophenyl ester **5** gave **Cy7⊂T6*** in 69% yield. Addition of this ligand to **c-P6** resulted in immediate and quantitative formation of the desired complex **Cy7⊂c-P6·T6***. The three rotaxanes, **4**, **Cy7⊂T6*** and **Cy7⊂c-P6·T6***, were thoroughly characterized by mass spectrometry and NMR spectroscopy, and their $^1$H NMR spectra were fully assigned using COSY and NOESY techniques, as detailed in the SI. In all of these compounds, the asymmetric environment of the α-CD makes all four methyl groups of the cyanine dye non-equivalent, which proves that the **Cy7** is inside the α-CD, as confirmed by the observation of NOEs from protons 3 and 5 of the α-CD to protons on the **Cy7** bridge. In **Cy7⊂c-P6·T6***, the asymmetry of the α-CD results in eight distinct porphyrin β-pyrrole doublets, and the pattern of NOEs demonstrates the close proximity of all three components (**c-P6**, CD and **Cy7**). The observation of NOEs from the α-CD-OMe groups to meso-aryl protons of the nanoring, and to the β-pyridyl resonances, is consistent with conformation **A**, in which the α-CD sits tightly within the nanoring.

Electronic coupling between **Cy7** and **c-P6** would be expected to result in changes in the intensity and/or wavelength of the optical transition of the **Cy7⊂c-P6·T6*** complex, compared with the transitions of free **Cy7⊂T6*** and **c-P6·T6***, as predicted for the bacterial LH1-RC supercomplex.[13] The absorption spectrum of the **Cy7⊂c-P6·T6*** complex is compared with the sum of the spectra of its components in Figure 2. The spectrum of the target complex is similar to the sum of the absorption spectra of its individual components. The three-peak structure of the Q-band is left mostly unaltered, suggesting that the spectral changes induced by a coherent excitonic coupling between the nanoring and the central dye may be too subtle to be detected experimentally. Nonetheless, as we can assess theoretically, this coupling is similar in magnitude to the excitonic interaction between LH1 and RC pigments[13] (see SI) and can influence strongly other photophysical properties of **Cy7⊂c-P6·T6**, such as energy transfer dynamics, as we will show later.

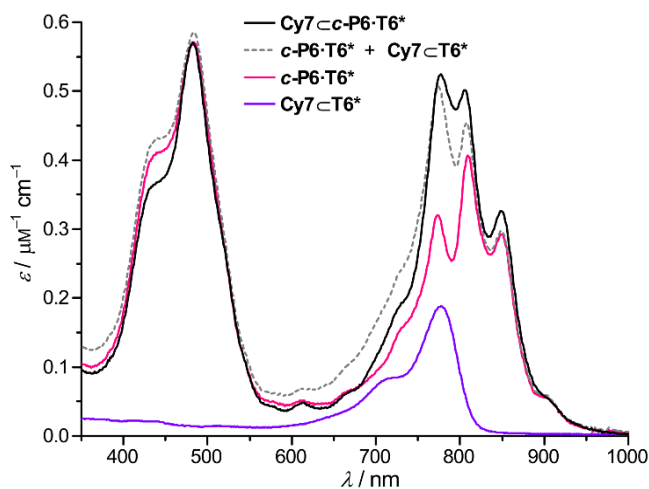

**Figure 2** Absorption spectra of **Cy7⊂c-P6·T6*** (black), **c-P6·T6*** (red), **Cy7⊂T6*** (purple), and the addition of absorption spectra of **c-P6·T6*** + **Cy7⊂T6*** (gray dashed line), in CH$_2$Cl$_2$ at 25 °C.

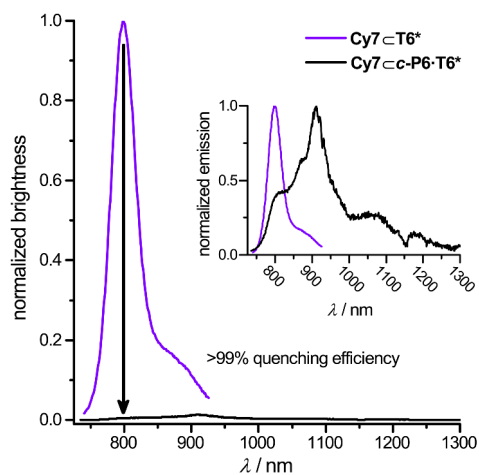

**Figure 3.** Normalized brightness of fluorescence spectra for **Cy7⊂T6*** and **Cy7⊂c-P6·T6*** excited at 725 nm. Normalized brightness is calculated by scaling the areas of the emission spectra by their quantum yields and to their molar absorption coefficients at 725 nm. The ratio of intensities at 798 nm is 0.0051. Inset shows the normalized emission spectra. Spectra recorded in CH$_2$Cl$_2$ at 25 °C.

The fluorescence behavior of **Cy7⊂*c*-P6·T6*** was examined to test how well it mimics the LH1-RC complexes. In Figure 3, the fluorescence spectra of **Cy7⊂*c*-P6·T6*** and **Cy7⊂T6*** are compared. Both samples were excited at 725 nm in $CH_2Cl_2$, and the intensities of the spectra are scaled so that their areas are proportional to the fluorescence quantum yields multiplied by the molar absorption coefficient at 725 nm, to give the normalized brightness. The cyanine dye rotaxane **Cy7⊂T6*** exhibits a bright emission band at 798 nm (quantum yield: $\Phi_f$ = 36%). In contrast, **Cy7⊂*c*-P6·T6*** is weakly fluorescent ($\Phi_f$ = 0.6%; $\lambda_{max}$ = 910 nm), and its brightness at 798 nm is 0.0051 (relative to 1.0 for **Cy7⊂T6***, both excited at 725 nm). Thus the presence of the surrounding nanoring quenches more than 99% of the emission from the cyanine core of **Cy7⊂*c*-P6·T6***. The emission spectra and quantum yields of **Cy7⊂*c*-P6·T6*** and ***c*-P6·T6*** are essentially the same (see SI, Figure S37), which implies that there is efficient energy transfer from **Cy7** to ***c*-P6** in **Cy7⊂*c*-P6·T6***. The fluorescence lifetime of **Cy7⊂T6*** is 0.97 ns (see SI), so the reduction in brightness by a factor of <99% indicates that energy transfer occurs on a time scale shorter than 10 ps. The excitation spectrum of **Cy7⊂*c*-P6·T6*** (measured at the emission maximum at 910 nm; SI, Figure S36) matches its absorption spectrum, indicating the energy absorbed at all wavelengths undergoes transfer and fluorescence with the same efficiency. The fact that the surrounding ring fluoresces, emulates analogous observations for the LH1-RC supramolecular complex.[14] Although the **Cy7** dye and the ***c*-P6** ring absorb light almost independently, as discussed above (Figure 2), the energy absorbed by the chromophores in both structures is transferred to the lowest energy excited state efficiently before fluorescence occurs. If the energy transfer were not efficient, emission would take place from all the transitions that absorb light and with a significant contribution from the **Cy7** dye, as it has a larger fluorescence quantum yield than the ring (see SI, Table S5). In contrast, we verify that emission occurs from this lowest energy state regardless of the excitation wavelength (see SI, Figure S36), in line with efficient exciton transfer dynamics enabling fluorescence after thermal equilibration to the lowest energy state of the full dye-ring complex. In order to clarify the mechanism that leads to the efficient energy transfer that we observe experimentally (Figure 3), we propose a microscopic model based on Davydov-Frenkel theory of molecular excitons[15] which describes the electronic transitions of the Q-band in terms of coupled transition dipoles associated to individual porphyrins (for more details, see SI, Section 7). This model reproduces the main structure (optical transitions and line-shapes) in the absorption spectrum of the nanoring (Figure 4a, see SI Section 7 for more details). Moreover, it predicts a small but persistent dipole redistribution from the 810 nm band of the nanoring to the central dye due to a small coherent dye-nanoring coupling (~25 $cm^{-1}$, Figure 4b), consistent with predictions for the LH1-RC supercomplex.[13] Although the coupling is too small to be observed directly in the experimental absorption spectra of Figure 2, it leads to very efficient energy transfer from the dye to the nanoring, typically within 3 ps (Figure 4c). This value perfectly matches our estimates based on fluorescence measurements, thus further validating our theoretical model.

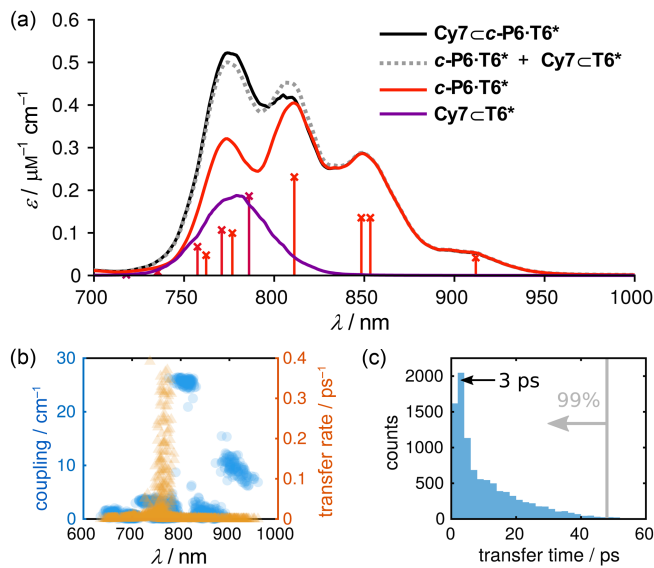

**Figure 4.** Theoretical description of absorption spectra and transfer dynamics using an excitonic model system (see SI for details). (a) theoretical absorption spectra of **Cy7⊂*c*-P6·T6*** (black), ***c*-P6·T6*** (red), **Cy7⊂T6*** (purple), and the addition of theoretical absorption spectra of ***c*-P6·T6*** + **Cy7⊂T6*** (gray dashed line). The vertical lines represent the stick spectrum of **Cy7⊂*c*-P6·T6***. (b) Coupling (blue circles) and transfer rate (orange triangles) between the dye and bands of the nanoring at different wavelengths. (c) Distribution of energy transfer times from **Cy7** to the nanoring, calculated within second order perturbation theory in the dye-nanoring coupling. The energy transfer takes less than 50 ps with 0.99 probability.

The **Cy7⊂*c*-P6·T6*** complex is a much simpler molecular system than the LH1-RC supercomplex, but it provides a model to study and engineer the underlying mechanism behind the energy transfer in complex natural light-harvesting systems. This work confirms that a special arrangement of dyes, similar to the LH1-RC complexes, with one dye at the center of a ring of other dyes, provides highly efficient energy transfer between the peripheral and central chromophores.

## Acknowledgements

This work was supported by the European Research Council via the Synergy grant BioQ (Grant no. 319130) and was made possible through the support of a grant from the John Templeton Foundation. The opinions expressed in this publication are those of the authors and do not necessarily reflect the views of the John Templeton Foundation. ATF thanks Christ Church, Oxford, for a Junior Research Fellowship.

**Keywords:** light harvesting • porphyrin • cyanine • rotaxane • excitonic coupling

**Entry for the Table of Contents**

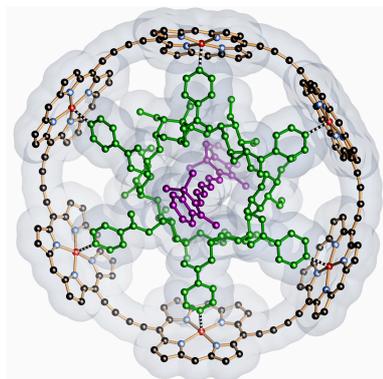

**Colored to the core:** Rotaxane formation has been used to lock a cyanine dye along the axis of a porphyrin nanoring, to form a complex that mimics the photophysical behavior of the light harvesting complex LH1 and the reaction center in purple photosynthetic bacteria. Energy is transferred rapidly between the axial dye and the surrounding nanoring.

Institute and/or researcher Twitter usernames: @HLAGroupOx and @qubit_ulm